\documentstyle[12pt]{article}
\newcommand{\LI}{\hbox to\hsize}
\newcommand{\LLI}[1]{\LI{#1\hss}} \newcommand{\RLI}[1]{\LI{\hss#1}}
\newcommand{\CLI}[1]{\LI{\hss#1\hss}}

\newcommand{\PM}[1]%
{\mbox{$m_{\rm #1}$}} %generic particle mass
\newcommand{\BEQ}{\begin{equation}}
\newcommand{\EEQ}{\end{equation}}
\newcommand{\lappr}{\mbox{$\stackrel{<}{\sim}$}} % less than, approximately
\newcommand{\ack}{\vskip6mm \LLI{\Large{\bf Acknowledgement}}
\vspace*{2mm}
\par
\noindent
\nopagebreak
This research has been supported in part by the U.S. Department of Energy
under Grant \#~DE--FG--02--85ER40211.} % Acknowledges grant support
\newcommand{\incircle}[1]{\mbox{{\hbox{$\bigcirc$}\kern-0.7em
\lower0.05ex\hbox{\mbox{{\scriptsize\rm #1}}}}}}
\newcommand{\eq}[1]{eq.~(\ref{#1})}
\newcommand{\ETC}{\mbox{\em etc.\/ }}
\newcommand{\VIZ}{\mbox{\em viz.\/ }}
\newcommand{\CF}{\mbox{\em cf.\/ }}
\newcommand{\IE}{\mbox{\em i.e. \/}}
\newcommand{\ETAL}{\mbox{\em et. al.\/ }}
\newcommand{\EG}{\mbox{\em e.g.\/ }}

\newcommand{\mathfig}[4]{%
\begin{figure}[#4] % position the figure by #4
\vspace*{6in}
\centerline{\hbox to 14in % center the file and set its width
 {\hskip5in {\special{eps:#2 x=14in}\hfil}}} % cut off a piece of the
\vspace*{-1.5in} % move up the the caption
\caption{#3}
\label{#1}
\end{figure}\vskip0.5in} % leave space between the caption and the next
\begin{document}
\RLI{JHU--TIPAC 940014}
\RLI{August 1994}
\vskip5mm
\begin{center}
{\Large\bf Localized Energy Deposition in Neutrino Telescopes:\\[1mm]
a Signature of ``New Physics''}\\[4mm]
G. Domokos and S. Kovesi--Domokos\\
INFN, Sezione di Firenze\\
Florence, Italy\\
and\\
The Henry A. Rowland Department of Physics and Astronomy\\
The Johns Hopkins University\\
Baltimore, MD 21218\footnote{Permanent address. E--mail:~SKD@JHUP.PHA.JHU.EDU}

\end{center}

{\small A class of phenomena either not contained in the
Standard Model or in its perturbative treatment
(multiple W/Z production, compositeness of quarks
and leptons) leads to a large energy deposition in
neutrino telescopes within a rather small volume. The
shape of the particle distribution arising from such phenomena is
estimated.} \vskip6mm

Neutrino telescopes may play a significant role in particle physics as well
as in neutrino astronomy. The basis for their value as particle
physics tools lies
in the fact that, according to a variety of estimates, \CF refs.~\cite{flux},
active galactic nuclei (AGN) as well as other point sources are likely to be
sources of ultra high energy (UHE) neutrinos, with energies up to,
perhaps, several TeV in the CMS in the interaction with nucleons
(energies of the order of a few times $10^{7}$GeV in the laboratory).
No currently functioning  or planned terrestrial device
can compete with such neutrino energies. If the detector is sufficiently
large, the limitation due to the low expected flux can be, in part,
overcome. Even though the subject of UHE neutrino emission from
AGNs and other point sources is at present a controversial one,
in a few years, there will be a sufficient amount of data from
neutrino telescopes available in order to confirm the existence
(or absence) of such neutrinos. What follows is contingent upon
the existence of UHE neutrinos emitted by {\em some} sources.

An obvious advantage of the experimentation with neutrino beams is
a relatively low background of ``mundane'' physics; hence, one expects
the study of neutrino induced reactions to serve as a useful tool
in looking for phenomena beyond the Standard Model, \CF~\cite{pylos92}.

We propose that if there exists another level of compositeness with a
characteristic energy scale of a few TeV, \cite{domonuss, prcomposite},
one is likely to discover
it in a neutrino telescope by means of a large energy release in a
small volume within or near the telescope. Likewise, if the conjectured
phenomenon of multiple W/Z production (with or without the violation of
B$+$L, see~\cite{thooft}) exists with a comparable cross
section~\cite{ringwald}, one will observe a somewhat similar phenomenon.
We conjecture that one will be able to distinguish between the two types
of phenomena experimentally; we return to the discussion of this topic
later.

The phenomena referred to above are conjectured to have a cross section
of the order of a few microbarns, \CF~\cite{prcomposite, ringwald}. In
addition, both processes are characterized by the production of a substantial
number ($5\lappr \langle N \rangle \lappr 30$) of energetic hadrons:
in the case of  composite models of quarks and leptons this was estimated in
ref.~\cite{prcomposite}; in the case of multiple gauge boson production, it
is a consequence of the fact that the weak gauge bosons decay into hadron
with a branching ratio of about 70\%, see~\cite{particledata}.

Next, we notice that in a ``typical'' neutrino telescope, for instance,
DUMAND or NESTOR,  the detector lies beneath about 4 km. of
water, corresponding to a thickness, $d\approx 4\times 10^{5} {\rm g/cm}^{2}$
at zero zenith angle.
At a non--vanishing zenith angle, $\zeta$, the absorber thickness
is given to a good approximation by the elementary formula:
\BEQ
t(\zeta)\approx R \left( \sqrt{\left(\cos \zeta\right)^{2}
+ 2d/R} - \cos \zeta \right),
\label{eq:thickness}
\EEQ
where $R$ is the radius of the Earth. (This formula is valid for
$ d\ll R$.)
The formula expressing the interaction mfp., $\lambda$,
in terms of
the cross section (valid to a good approximation for {\em small}
cross sections~\cite{darkmatter}) is:
\BEQ
\lambda \approx \frac{1670}{\sigma},
\label{eq:mfp}
\EEQ
where $\sigma$ is measured in millibarns and $\lambda$ in
${\rm g/cm}^{2}$. Thus, in the range of zenith
angles, $0\lappr \zeta \lappr \pi /2$, the mfp. varies between
$d$ and about $2\times 10^{7}{\rm g/cm}^{2}$. Correspondingly,
if the cross section of the process to be observed satisfies the
approximate inequality,
\[
7 \times 10^{-5}{\rm mb}\quad \lappr \quad \sigma
\quad \lappr \quad 4\times 10^{-3}{\rm mb},
\]
the first interaction,
on the average, takes place in or near the
neutrino telescope. Smaller cross sections --- all the way down to the
Standard Model weak interaction cross section --- can be explored in a
similar way. However, the estimate of the mfp. quoted above is no longer
reliable, since the amount of matter between the detector and the incident
beam depends strongly on the geology of the environment of the telescope.

Due to the fact that the initial interaction takes place in or near
the detector, the produced hadrons initiate a mixed,
hadronic~--~electromagnetic cascade right  around the detector.
Due to the high density and low average nuclear charge ($Z_{eff} \approx 3.3$)
of the environment in which the cascade
evolves, such water showers have some  peculiar features both with respect to
cascades developing in air and in absorbers like Pb.
(For all practical purposes, one can take $\rho \approx
1$ even at a depth of 4 km due to the low compressibility of water.)

We now turn to estimating some important characteristics of these showers
in order to be able to recognize them in a neutrino telescope.

Let us notice the following qualitative features of a water shower.
\begin{itemize}
\item The shower is nucleon poor. In fact, baryon pair production is rare
in any hadronic interaction; moreover, due to the low effective atomic
number of the environment, ($A_{eff} = 6$) the probability of knocking out
a substantial number of the target nucleons in any given interaction is
small.
\item With the exception of particles containing c, b and t quarks produced
in any of the
interactions, charged particles interact rather than decay at all energies
of interest. In fact, for charged pions (the most copiously produced
charged hadrons), the interction mfp. is about 78 ${\rm g/cm}^{2}$,
see \EG~\cite{particledata}, whereas the decay length in water is
$\approx 800 E/m_{{\rm \pi}} \quad {\rm g/cm}^{2}$. Neutral pions still decay
at
most energies of interest: with a decay length, $c\tau \approx
25 {\rm nm}$, the interaction and decay mfp. in water become
roughly equal at $E_{\pi^{0}}\approx 4$PeV.
\end{itemize}

We calculate the cascade development making the following approximations.
\begin{enumerate}
\item We neglect the production of all particles except pions. We further
assume
that pions of all charges are produced at the same rate.
\item We completely neglect the decay of charged pions and the interaction
of neutral ones.
\item We neglect photoproduction of pions.
\item We compute the cascade development in Approximation A both
for the hadronic and electromagnetic components. (The cascade is
computed in the diffusion approximation; lateral development is
neglected. In the case of the hadronic component, processes like
pion capture, nuclear breakup, \ETC are neglected. Likewise, in the
development of the electromagnetic component, Compton scattering and ionization
are neglected.)
\item We assume that, apart from the initial interaction, the
evolution of the cascade takes place according to the Standard Model.
This is justified by the circumstance that --- as described in
the references quoted --- the cross section of either process considered
here is expected to rise rapidly around its characteristic energy and
then go into saturation.
\end{enumerate}
As discussed elsewhere, \CF~\cite{pylos93}, this approximation should be
adequate for gaining an understanding of the main qualitative features
of the cascade. Due to the fact that one's understanding of the initial
interaction is very sketchy at best, a more accurate computation of a
cascade without knowing the details of the process initiating it is
not warranted.

With these premises, we now turn the computation. Due to the fact that
photoproduction is neglected, the hadronic and electromagnetic parts
of the cascade can be calculated in separate steps. The hadronic component
evolves autonomously; neutral pion decay merely acts as a source for
the electromagnetic component.

In order to compute the hadronic component, we write down the standard
transport equation for a {\em single component cascade}, since there are only
pions present. We have:
\BEQ
\frac{\partial H(E,x)}{\partial x} = - H(E,x)
+ \int_{E}^{\infty} \frac{dE'}{E} F\left(E, E'\right) H(E',x).
\label{eq:transport}
\EEQ
Here, $x$ stands for the thickness of the absorber measured in units of
the interaction mfp. and $H(E,x)$ is the differential distribution of the
pions.
The fragmentation function, $F$, is taken to be of the form:
\BEQ
F(E, E') = \Theta \left(E' - E_{crit}\right)f(z),
\EEQ
where $z$ is a Feynman scaling variable, $z=E/E'$; the energy
$E_{crit}$ is the energy below which the produced pions have such a low
energy that dissipative processes (nuclear breakup, \ETC )
begin to compete successfully with particle production. We found that
taking $E_{crit}\approx 500$GeV (corresponding to $\sqrt{s} \approx
30$GeV) is an adequate choice.

The scaling function, $f(z)$, can be reasonably well approximated
by an expression of the form:
\BEQ
f(z) =  C z^{-0.9} \left( 1 -z\right)^{3}\Theta\left( z - z_{min}\right).
\EEQ
The parameters $C$ and $z_{min}$ are determined from the conditions of
energy conservation and from the value of the charged particle multiplicity
(assumed to be energy independent), \CF~\cite{pylos93}. We found
$C\approx 0.45$ and $z_{min} \approx 0.008$. Apart from the presence
of a critical energy, $E_{crit}$, all violations of Feynman scaling
are neglected.

In order to solve \eq{eq:transport}, one uses a combination of analytical
and numerical methods as described in \cite{pylos93}; this assures a much
higher accuracy than the asymptotic methods (saddle point approximation,
\ETC) used before digital computers appeared. In essence, the method consists
of generating an iterative solution to the transport equation. The resulting
integrals are then evaluated numerically. The computational time
rquired is a small fraction of what would be needed for a full scale
Monte Carlo calculation. (As discussed before, any gain in accuracy
by using a Monte Carlo simulation is a purely illusory one.)

The transport equation is solved with an initial condition representing the
``new physics''. Due to the theoretical uncertainties, we chose a very
simple model of the energy distribution, \VIZ we distributed the primary
energy equally among the secondaries emerging from the first interaction.
In a different context, when the characteristics of the first interaction
were known, it was checked that this approximation is adequate if the
required accuracy is not higher than about 50\%, \cite{darkmatter};
hence, it should be adequate here. We put for a fixed primary energy:
\BEQ
N(E, 0) = N_0 \delta \left( E - E_0\right),
\label{eq:inihadron}
\EEQ
where $E_0$ and $N_0$ are the primary energy and initial multiplicity.

One cannot find simple analytic expressions for the solution of
\eq{eq:transport}. In  Figures 1 and 2 we display the integral
distribution of charged hadrons,
\[
N_(>E, x) = \int_{E}^{E_0} dE' N(E',x)
\]
for $E= 10$Gev and $E= 100$GeV, respectively; the primary energy
and initial multiplicity
chosen were $E_{0} = 10$PeV and $N_{0}=20$. (The value chosen for the initial
multiplicity is not critical, since the solution depends linearly on it.)

One notices that --- as expected --- the integral distribution for a lower
$E$ is
broader in $x$ than the one for a high energy.
In essence, this is a consequence
of energy conservation: particles of lower energy occur further downstream
than ones of high energy.

The transport equation for the electromagnetic component ($e^{\pm},
\gamma$) is handled in a similar fashion. Apart from the very beginning of an
electromagnetic cascade, the number of electrons, positrons and photons
are almost equal to each other at the relevant energies. Consequently,
they can be adequately described by a transport equation of the
form,
\BEQ
\frac{\partial \Gamma(E, x_{R})}{\partial x_{R}} =
- \Gamma (E, x_{R}) + \int_{E}^{\infty}
\frac{dE'}{E} \phi \left(\frac{E}{E'}\right) \Gamma (E', x_{R})
+ S(E, x_{R}).
\label{eq:elmag}
\EEQ

In the last equation, $x_{R}$ stands for the absorber thickness measured in
units of the
radiation length and  $\Gamma$ stands for the distribution of photons;
as asserted, at all distances but the very smallest ones, one can take
the distributions of the photons and charged leptons to be the same.
The function $\phi$ is the fragmentation function in the traditional
Heitler--J\'{a}nossy approximation, \CF~\cite{rossi}. (The fact that
putting $\phi(E,E') \approx \delta (E-E'/2)$ is an
acceptable  approximation
explains, in essence, why the electron and
photon distributions are about the same.) Finally, $S(E,x_{R})$ is the
source of the electromagnetic component. To a fair approximation,
\[
S\left( E, x_{R}\right) = H\left( 2E, x_{R}\right).
\]
This is due to the fact that photons almost exclusively arise
from the decay of neutral pions; furthermore, at high energies, the photons
share the energy of the decaying $\pi^{0}$ equally, and, finally, the
number of photons from $\pi^{0}$ decay is, to a good approximation,
equal to the number of charged pions.

The solution of the transport equation \eq{eq:elmag} can be obtained in the
same way as described in connection with the transport equation for hadrons.
In Fig. 3 we display the integral distribution of the electromagnetic
component for the same
initial conditions as used in computing the hadron distribution. For the
ease of comparison, the hadronic and electromagnetic components are
plotted using the same unit of length (meters), assuming a constant
density, $\rho \approx 1$.

The distributions shown in Figures 1 through 3 are {\em typical\/}: variations
in the primary energy by factors of about 100 change the total number of
particles, but they, in essence, leave the profiles of both the hadronic
and leptonic components unchanged.

One  concludes that events of the type described above
are characterized by a large amount of energy released around the
detector within a rather small volume. The
characteristic longitudinal distances are about 10 meters or less
and, due to the
high energies involved, the transverse size is considerably smaller.
The energy released in this volume is close to
the energy of the incident neutrino. We believe
that such a signature is practically free of background. However, this
question cannot be answered without detailed and detector dependent
simulations.

Some potential sources of background  one can think of are:
\begin{itemize}
\item Normal neutrino interactions occurring close to the detector,
arising from neutrinos incident at a large zenith angle ($\zeta
\approx \pi/2$, say, due to fluctuations in the occurrence of the
first interaction.
\item One half of the Learned--Pakvasa process~\cite{learned}, \IE
either the production or the decay of a $\tau$ occurring near the detector,
the other half being sufficiently far away so as to escape detection.
\item Muons of energies in the 10 PeV range (or larger) passing through the
detector.
Such muons lose energy at the rate of
about 0.5 TeV/m and thus, they may deposit a substantial amount of energy
in the detector within a short distance.
\end{itemize}

While a quantitative answer requires a detailed study of the characteristics
of each type of event mentioned above and a careful simulation
of their appearance, one is under the impression that the signal we
described here stands out by the large {\em energy density}
deposited in the detector.

One should also study in more detail the possibility of distinguishing
between the events generated by ``composite neutrinos''
(see~\cite{domonuss}) and multiple gauge boson production. At this point
we can make a few qualitative conjectures only. In the case of the
multiple production of gauge bosons, about 70\% of the W/Z bosons
decays hadronically, and the remainder decays into  ${\rm e, \mu ,
\tau}$, with a branching ratio of about 10\% each. Thus, in such an
event, the electromagnetic component of the cascade starts early,
it being fed initially by the prompt electrons coming from
the decay of the gauge bosons. It is not clear at this point whether
such an effect can be observed in the presence of the large hadronic
component. Probably, one has a better chance of observing the
muons emerging from the decay as it was pointed out by
Ringwald,~\cite{ringwald}.

By contrast, the ``composite neutrino'' scenario will, probably,
lead to a smaller multiplicity of muons (probably, just one hard
muon at the vertex where the neutrino breaks up into its
``preons''~\cite{mrenna}). Hence, the observable to be watched in this
respect is the number and energy distribution of prompt muons.

In a recent publication~\cite{kamiokande}, the KAMIOKANDE
collaboration described a somewhat unusual occurrence. Apparently,
a particle penetrated about 3 km of rock\footnote{A.K. Mann,
private communication} corresponding to an absorber depth of
perhaps $10^{6} {\rm g/cm}^{2}$ after which it interacted
in the neighborhood of the detector. As a result of the interaction,
a large amount of energy was deposited in the detector: a substantial
number of PMTs having been saturated. Unless the event is due to
a very rare fluctuation, the interaction cross section of the
primary was about $10^{-3}$mb or so. It is amusing to speculate that
thereby  the first event of the type described in this letter
may have been observed.
\ack
\vskip1mm This work was completed during the authors' visit at the
University of Florence, Summer 1994. We thank Roberto Casalbuoni for
the  hospitality extended to us
and the National Institute for Nuclear Physics (INFN)
for partial financial support.

\newpage
\CLI{\Large Figure Captions}\vskip6mm
Fig.~1\hskip2mm Integral spectrum of hadrons of energy $E>10$GeV.
The primary energy
of the first interaction is assumed to be $E_{0}=10^{7}$GeV, initial
multiplicity, $N_{0}=20$. The hadronic cascade shown is the one
generated by an ``average hadron'' emerging from the first interaction.
\vskip4mm
Fig.~2\hskip2mm Same as Fig.1, for hadrons of energy $E>100$GeV.
\vskip4mm
Fig.~3\hskip2mm The electromagnetic component of the shower generated by
an ``average hadron'' emerging from the first interaction; leptons
(${\rm e},{\rm \bar{e}},\gamma$) with energy, $E>10$GeV. For comparison,
the hadronic component is plotted by a dotted line.

\begin{thebibliography}{99}
\bibitem{flux} The subject has been thoroughly reviewed in:
``Proceedings of the Workshop on High Energy Neutrino
Astrophysics''; V.J. Stenger, J.G. Learned, S. Pakvasa and
X. Tata, editors. World Scientific Publishing Company,
Singapore (1992). For recent work, \CF A.P. Szabo and R.J. Protheroe,
University of Adelaide preprint ADP-AT-94-4.
%Protheroe and Szabo, Begelman, etc.
\bibitem{pylos92} G. Domokos anD S. Kovesi--Domokos, in Proceedings
of the Second NESTOR Workshop, L.K. Resvanis, editor. University of
Athens Press, (1993).
\bibitem{domonuss} G. Domokos and S. Nussinov, Phys. Letters
{\bf B87} (1987) 372 %Domokos & Nussinov
\bibitem{prcomposite}G. Domokos and S. Kovesi--Domokos,
Phys, Rev. {\bf D 38} (1988) 2833 % P.R. D article on compositeness
\bibitem{thooft} G. `t Hooft, Phys. Rev. {\bf D 14} (1976) 3432;
{\em ibid.\/} {\bf D 18} (1978) 2199 % B+L violation by instantons.
\bibitem{ringwald} A. Ringwald, in ``Proceedings of the $17^{\rm th}$
Johns Hopkins Workshop on Current Problems in Particle Theory'';
Z. Horv\'{a}th and A. Patk\'{o}s, editors. World Scientific Publishing
Company, Singapore (1994)
\bibitem{darkmatter} G. Domokos, B. Elliott and S. Kovesi--Domokos,
Jour. Phys. G {\em 19} (1992) 899 % J.Phys. G
\bibitem{particledata} Particle Data Group, ``Review of Particle Properties'',
Phys. Rev. {\bf D 45} (1992) Number 11, Part II
\bibitem{pylos93} G. Domokos and S. Kovesi--Domokos, in ``Proceedings of the
Third NESTOR Workshop'', L.K. Resvanis, editor. Univerity of Athens Press,
to appear.
\bibitem{rossi}B. Rossi, High Energy Particles. Prentice Hall, New
York, (1952).
\bibitem{mrenna} S. Mrenna, Phys. Rev. {\bf D45} (1992)  2371
\bibitem{kamiokande} KAMIOKANDE collaboration, K.S. Hirata \ETAL,
Phys. Rev. {\bf D 45} (1993) 3345.
\bibitem{learned} J.G. Learned and S. Pakvasa, Preprint DUMAND--3--94
(1994).
\end{thebibliography}
\end{document}